\documentclass[english,aps,prd,floats,floatfix,nofootinbib]{revtex4}
\usepackage{natbib}
\usepackage{times}
\usepackage{float}
\usepackage{amsmath}
\usepackage{graphicx}
\usepackage{verbatim}
\usepackage{amssymb}
\usepackage{slashed}
\usepackage{color}
\usepackage[dvipsnames]{xcolor}
\usepackage{soul}
\usepackage[normalem]{ulem}

\def\slashp{p \!\!\! \slash}

\newcommand{\eR}{\epsilon_R}

\begin{document}
\title{New physics in the kinematic distributions of $\bar B\to D^{(*)}\tau^-(\to\ell^-\bar\nu_\ell\nu_\tau)\bar\nu_\tau$}
\author{Rodrigo Alonso$^1$, Andrew Kobach$^1$ and Jorge Martin Camalich$^{2}$}
\affiliation{
$^1$Dept. Physics, University of California, San Diego, 9500 Gilman Drive, La Jolla, CA 92093-0319, USA\\
$^2$PRISMA Cluster of Excellence \& Institut f\"ur Kernphysik, Johannes Gutenberg Universit\"at Mainz, 55128 Mainz, Germany}

\begin{abstract}
We investigate the experimentally-accessible kinematic distributions of the $\bar B\to D^{(*)}\tau^-(\to\ell^-\bar\nu_\ell\nu_\tau)\bar\nu_\tau$ decays. 
Specifically, we study the decay rates as functions of the $B\to D^{(*)}$ transferred squared momentum, the energy of the final charged lepton and 
the angle of its 3-momentum with respect to the 3-momentum of the recoiling $D^{(*)}$. The angular distribution allows to introduce new observables, 
like a forward-backward asymmetry, 
which are complementary to the total rates. We present analytic formulas for the observable 3-fold 5-body differential decay rates, 
study the predictions in the Standard Model and investigate the effects in different new-physics scenarios that we characterize using an 
effective field theory framework.
\end{abstract}
\maketitle

\section{Introduction}

The $\bar B\to D^{(*)}\tau^-\bar\nu$ decays manifest some of the most prominent anomalies in low-energy
flavor observables. Significant enhancements of the rates with respect to the Standard Model (SM) predictions are observed 
in the two decay channels ($D$ and $D^*$) and by three different experiments, BaBar~\cite{Lees:2012xj,Lees:2013uzd}, 
Belle~\cite{Huschle:2015rga} and LHCb~\cite{Aaij:2015yra}. The anomalies appear in the ratios 
$R_{D^{(*)}}=\mathcal{B}(\bar B\to D^{(*)}\tau^-\bar\nu)/\mathcal{B}(\bar B\to D^{(*)}\ell^-\bar\nu)$, with $\ell=e,\,\mu$,
where many of the experimental and theoretical uncertainties cancel. 

The theoretical predictions of the $B\to D^{(*)}\tau^-\bar\nu$ rates in the SM are very accurate and rely on parametrizations of the form factors in an expansion about
the heavy-quark limit including up to $\mathcal{O}(1/m_Q)$ corrections and constraints
from unitarity~\cite{deRafael:1993ib,Boyd:1995sq,Caprini:1997mu}. The normalization at zero-recoil (which includes $|V_{cb}|$) and kinematic
dependence of the form factors in these parametrizations are fitted to the total width and spectra of the decays into the
light leptons~\cite{Amhis:2014hma}. For the decays into $\tau$ leptons the amplitude becomes sensitive to scalar form factors for which calculations 
in Lattice QCD (LQCD) become necessary~\cite{Becirevic:2012jf,Lattice:2015rga,Na:2015kha,Du:2015tda}. 

The discrepancy between experiment and the SM is at the level of $4\sigma$, and it can be explained with 
new physics (NP)~\cite{Fajfer:2012vx,Crivellin:2012ye,Celis:2012dk,Becirevic:2012jf,Datta:2012qk,Tanaka:2012nw,Ko:2012sv,Sakaki:2013bfa,Duraisamy:2013kcw,Biancofiore:2013ki,Sakaki:2014sea,
Alonso:2015sja,Freytsis:2015qca,Barbieri:2015yvd,Crivellin:2015hha,Calibbi:2015kma,Fajfer:2015ycq,Bhattacharya:2015ida,Bauer:2015knc}.
These analyses of the data are most fruitful when casted model-independently in an effective field theoretical framework. The results obtained in this approach can 
then be used as input to determine which models could explain the putative effect. In addition to $R_{D^{(*)}}$, the spectra in $q^2$ of the rates have also been 
reported by BaBar and Belle, which is useful to discriminate among the different possible NP contributions~\cite{Lees:2013uzd,Freytsis:2015qca}. 

However, none of the phenomenological analyses study the full kinematic distributions of the 5-body decays (or 6-body if we include the decay of the $D^*$), 
despite the fact that the experiments exploit them to extract the signal from background, mainly through Monte-Carlo simulation.
Besides the dependence of the rate on $q^2$ and the final charged lepton energy, one can 
also study the dependence on the angle that the 3-momentum of this final lepton forms with the recoil direction of the $D^{(*)}$~\cite{Nierste:2008qe,Tanaka:2010se,Hagiwara:2014tsa}. The
expected increase of statistics at the LHCb~\cite{Aaij:2015yra} and Belle2~\cite{Bevan:2014iga} encourages the exploration of the
the discriminating power of these distributions from a theoretical point
of view. Those in $q^2$ and $E_\ell$ are being used by BaBar and Belle, while the angular distribution enters indirectly in the dependence on the invariant missing mass of
the decay. Casting  the differential decay rate as an angular distribution offers a new method to not only discriminate among NP but also 
to increase the efficiency in the selection of the $\bar B\to D^{(*)}\tau^-\bar\nu$ signal events over the normalization mode,
$\bar B\to D^{(*)}\ell^-\bar\nu$. In the following, we investigate the experimentally-accessible kinematic distributions of the 
$\bar B\to D^{(*)}\tau^-(\to\ell^-\bar\nu_\ell\nu_\tau)\bar\nu_\tau$ decays, in generic scenarios of NP described 
using an effective-field theoretical framework.  

\section{The $\bar B\to D^{(*)}\tau^-\bar\nu$ differential decay rates}
\label{sec:3body}
\subsection{The low-energy effective Lagrangian}

The low-scale  $O(m_b)$  effective Lagrangian for semileptonic
$b \to c$ transitions is~\cite{Cirigliano:2009wk}:
\begin{eqnarray}
{\cal L}_{\rm eff} 
&=&
- \frac{G_F^{(0)} V_{cb}\eta_{\rm ew}}{\sqrt{2}} \,\sum_{\ell=e,\mu,\tau}
\Bigg[
\Big(1 + \epsilon_L^{\ell}  \Big) \bar{\ell}  \gamma_\mu  (1 - \gamma_5)   \nu_{\ell} \cdot \bar{c}   \gamma^\mu (1 - \gamma_5 ) b
+  \eR^{\ell}  \   \bar{\ell} \gamma_\mu (1 - \gamma_5)  \nu_\ell    \ \bar{c} \gamma^\mu (1 + \gamma_5) b\nonumber\\
&+& \bar{\ell}  (1 - \gamma_5) \nu_{\ell}
\cdot \bar{c}  \Big[  \epsilon_S^{\ell}  -   \epsilon_P^{\ell} \gamma_5 \Big]  b
+ \epsilon_T^{\ell}    \,   \bar{\ell}   \sigma_{\mu \nu} (1 - \gamma_5) \nu_{\ell}    \cdot  \bar{c}   \sigma^{\mu \nu} (1 - \gamma_5) b
\Bigg]+{\rm h.c.}, 
\label{eq:leff1} 
\end{eqnarray}
where we use  $\sigma^{\mu \nu} =i[\gamma^\mu, \gamma^\nu]/2$, $G_F^{(0)}\equiv \sqrt{2}g^2/(8 M_W^2)$
is the tree-level definition of the Fermi constant, and  $\eta_{\rm ew}=1.006$ encodes universal
short-distance electroweak corrections to the SM contribution~\cite{Sirlin:1981ie} (we neglect similar corrections to the NP
contributions).
The magnitude of the $\epsilon_i^{\ell}$ coefficients is set by $ v^2/\Lambda^2$ where $\Lambda$ is the NP scale
so that in the SM they vanish leaving the well-known
$(V-A)\times(V-A)$ structure generated by the exchange of a $W$ boson. The  $\epsilon_i^{\ell}$ coefficients
can display a scale dependence (together with the corresponding hadronic matrix elements). We have assumed that potential
right-handed neutrino fields (sterile with respect to the SM gauge group) are heavy compared
to the low-energy scale and have been integrated out of the low-energy effective theory.\footnote{Effective operators containing right-handed
neutrinos do not interfere with the SM amplitude and therefore contribute at $\mathcal{O}(\epsilon_i^2)$
to the decay rate.} If the NP is coming from dynamics at $\Lambda\gg v$ and electroweak symmetry breaking is linearly realized, then
 an effective $SU(2)_L\times U(1)_Y$ invariant effective theory applies~\cite{Buchmuller:1985jz,Cirigliano:2009wk}. A non-trivial
consequence of this is that, at leading order in the matching between the high- and low-energy theories~\cite{Alonso:2015sja}:
\begin{equation}
\epsilon_R^{\ell}=\epsilon_R^{\ell^\prime}+\mathcal{O}(v^4/\Lambda^4)\equiv \epsilon_R.\label{eq:RHCuniversal} 
\end{equation}
Therefore, any potential NP signal manifesting in $\epsilon_R$ will cancel to a large extent in the ratios $R_{D^{(*)}}$. 
Searches for this type of contributions can be done independently using the $\bar B\to D^{(*)}\ell^-\bar\nu_\ell$ decays~\cite{Becirevic:2016hea}.

\subsection{Form factors}

The hadronic matrix elements in the amplitudes derived from the effective Lagrangian in eq.~(\ref{eq:leff1}) are parametrized in terms of form factors,
\begin{eqnarray}
\langle D\left(k\right)  |\bar c\gamma^{\mu}b|\bar B\left(p\right)
\rangle&=& (p+k) ^\mu f_+(q^2) +(p-k) ^\mu f_-(q^2), \label{eq:ffSMD}\\
\langle D^{*}\left(k,\,\epsilon\right)  |\bar c\gamma^{\mu}b|\bar B\left(  p\right)\rangle&=& 
\frac{2\,i\,V(q^2)}{m_B+m_{D^*}}\varepsilon_{\mu\nu\alpha\beta}\epsilon^{*\nu}k^\alpha p^\beta,\label{eq:ffSMDsV}\\
\langle D^{*}\left(k,\,\epsilon\right)  |\bar c\gamma^{\mu}\gamma_5\,b|\bar B\left(p\right)\rangle&=&
2m_{D^*}A_0(q^2)\frac{\epsilon^*\cdot q}{q^2}q_\mu+(m_B+m_{D^*})A_1(q^2)\left(\epsilon^*_\mu-\frac{\epsilon^*\cdot q}{q^2}q_\mu\right)\nonumber\\
&-&A_2(q^2)\frac{\epsilon^*\cdot q}{m_B+m_{D^*}}\left(\left(p+k\right)_\mu-\frac{m_B^2-m_{D^*}^2}{q^2}q_\mu\right),\label{eq:ffSMDsA}
\end{eqnarray}
where $q=p-k$, $\varepsilon_{0123}=1$. The $f_-(q^2)$ can be written in terms of $f_+(q^2)$ and the scalar form factor
$f_0(q^2)$ using the conservation of the vector current in QCD,
\begin{eqnarray}
&&\langle D\left(k\right) |\bar{c}b|\bar B\left(p\right)
\rangle= \frac{m_B^2-m_D^2}{m_b-m_c}f_0(q^2),\nonumber\\
&&f_0(q^2)=f_+(q^2)+\frac{q^2}{m_B^2-m_D^2}f_-(q^2),\label{eq:ffNPDSc}
\label{eq:f0} 
\end{eqnarray}
with $f_0(0)=f_+(0)$. The pseudoscalar form factor for the $D^*$ channel can be related using partial conservation of the axial current:
\begin{eqnarray}
\langle D^{*}\left(k,\,\epsilon\right) |\bar c\gamma_5 b|\bar B\left(p\right)\rangle=\frac{2m_{D^*}}{m_b+m_c}A_0(q^2)\epsilon^*\cdot q\label{eq:ffNPDsPsSc}.
\end{eqnarray}
The matrix elements of the tensor operators are parameterized as:
\begin{eqnarray}
\langle D\left(k\right)|\bar c\sigma_{\mu\nu}b|\bar B\left(p\right)\rangle&=&\frac{2if_T(q^2)}{m_B+m_{D}}\left(k_\mu p_\nu-p_\mu k_\nu\right),\label{eq:ffNPDTens}\\
\langle D^{*}\left(k,\epsilon\right)|\bar c\sigma_{\mu\nu}b|\bar B\left(p\right)\rangle&=&\frac{\epsilon^*\cdot q}{(m_B+m_{D^*})^2}T_0(q^2)\varepsilon_{\mu\nu\alpha\beta}p^{\alpha}k^\beta
+T_1(q^2)\varepsilon_{\mu\nu\alpha\beta}p^\alpha\epsilon^{*\beta}+T_2(q^2)\varepsilon_{\mu\nu\alpha\beta}k^\alpha\epsilon^{*\beta}.\label{eq:ffNPDsTens}
\end{eqnarray}
These can be related to:
\begin{eqnarray}
\langle D\left(k\right)|\bar c\sigma_{\mu\nu}\gamma_5\,b|\bar B\left(p\right)\rangle&=&\frac{2f_T(q^2)}{m_B+m_{D}}\varepsilon_{\mu\nu\alpha\beta}k^\alpha p^\beta,\label{eq:ffNPDTens5}\\
\langle D^{*}\left(k,\epsilon\right)|\bar c\sigma_{\mu\nu}\gamma_5\,b|\bar B\left(p\right)\rangle&=&\frac{i\,\epsilon^*\cdot q}{(m_B+m_{D^*})^2}T_0(q^2)\left(p_\mu k_\nu-k_\mu p_\nu\right)
+i\,T_1(q^2)\left(p_\mu\epsilon^*_\nu-\epsilon^*_\mu p_\nu\right)\nonumber\\
&+&i\,T_2(q^2)\left(k_\mu\epsilon^*_\nu-\epsilon^*_\mu k_\nu\right),\label{eq:ffNPDsTens5}
\end{eqnarray}
through the relation $\sigma_{\mu\nu}\gamma_5=-i/2\,\varepsilon_{\mu\nu\alpha\beta}\sigma^{\alpha\beta}$.

\subsubsection{Numerical implementation}

The $q^2$-dependence of some of the form factors can be extracted experimentally analyzing the spectra of 
the $\bar B\to D^{(*)}\ell^-\bar \nu_\ell$ decays while a normalization factor -e.g. values of a form factor at $q^2=q^2_{max}$-
must be calculated using nonperturbative methods in order to extract $|V_{cb}|$ from the total rates. 
A particularly convenient parametrization is obtained using dispersion relations in QCD 
and heavy-quark effective field theory (HQET)~\cite{deRafael:1993ib,Boyd:1995sq,Caprini:1997mu}. In this parametrization the dependence on $q^2$ appears
through the product of the heavy-meson velocities,
\begin{equation}
w=v_{D^{(*)}}\cdot v_B=\frac{m_B^2+m_{D^{(*)}}^2-q^2}{2m_B m_{D^{(*)}}},
\end{equation}
where $q^2_{max}$ corresponds to $w_{min}=1$, or for a variable related by a conformal mapping which optimizes the convergence of a
Taylor expansion ($z$-expansions). In the conventional parametrization of ref.~\cite{Caprini:1997mu}, the $BD$ vector form factor is:
\begin{eqnarray}
f_{+}(w)=\frac{V_1(w)}{r_D},\hspace{0.5cm}V_1(w)=V_1(1)\left[1-8\rho_D^2z+(51\rho_D^2-10)z^2-(252\rho_D^2-84)z^3\right],\label{eq:ffparfp} 
\end{eqnarray}
where $r_D=2\sqrt{m_B m_D}/(m_B+m_D)$, $z=(\sqrt{w+1}-\sqrt{2})/\sqrt{w+1}+\sqrt{2})$ and $\rho_D^2$ is extracted from 
data. The scalar and tensor form factors are not measured and one needs to use HQET relations or LQCD. For $f_0$ one can 
use the expression derived in HQET~\cite{Caprini:1997mu,Tanaka:2010se}, but we rather implement the recent results obtained
in the lattice~\cite{Lattice:2015rga,Na:2015kha,Du:2015tda}, which are provided in terms of the $z$-expansion. For definiteness we use
the results of the HPQCD collaboration presented in ref.~\cite{Na:2015kha}:
\begin{equation}
f_0(w)=\frac{1}{P_0}\sum_{k=0}^2a_k^{(0)}z^{k}, \label{eq:ffparf0} 
\end{equation}
with $P_0=1-q^2(w)/M_0^2$. For $f_T$ there are no such calculations and we employ a relation that holds in the heavy-quark limit
and at leading order in $\alpha_s$:
\begin{equation}
f_T(w)=f_+(w)+\mathcal{O}(\Lambda/m_Q).\label{eq:ffparfT}
\end{equation}

\begin{table}[h]
\centering
\caption{Values for the form factor parameters employed in this work. The values for 
$\eta_{\rm ew}|V_{cb}|V_1(1)$, $\rho_D^2$, $\eta_{\rm ew}|V_{cb}|h_{A_1}(1)$, $\rho_{D^*}^2$, $R_1(1)$, $R_2(1)$ 
and their statistical correlations are obtained from the HFAG global fits to the $\bar B\to D^{(*)}\ell^-\bar\nu$ data~\cite{Amhis:2014hma}. Those 
for $V_1(1)$, $M_0$, $a_i ^{(0)}$ and their correlations are obtained from~\cite{Na:2015kha}, $h_{A_1}(1)$ from~\cite{Bailey:2014tva} and
$R_0(1)$ from ref.~\cite{Fajfer:2012vx}. \label{tab:FFparam}}
\begin{tabular}{|c|c|}
\hline
$BD$&$BD^*$\\
\hline
$\eta_{\rm ew}|V_{cb}|V_1(1)=42.65(1.53)\times10^{-3}$&$\eta_{\rm ew}|V_{cb}|h_{A_1}(1)=35.81(0.45)\times10^{-3}$\\
$\rho_D^2=1.185(54)$&$\rho_{D^*}^2=1.207(26)$\\
$V_1(1)=1.035(40)$&$h_{A_1}(1)=0.906(13)$\\
$M_0=6.420(9)$ GeV&$R_1(1)=1.406(33)$\\
$a_0^{(0)}=0.647(29)$&$R_2(1)=0.853(20)$\\
$a_1^{(0)}=0.27(30)$&$R_0(1)=1.14(10)$\\
$a_2^{(0)}=-0.09(2.24)$&\\
\hline
$C(\eta_{\rm ew}|V_{cb}|V_1(1),\rho_D^2)=0.824$&$C(\eta_{\rm ew}|V_{cb}|h_{A_1}(1),\rho_{D^*}^2)=0.323$\\
$C(a_{0}^{(0)},a_{1}^{(0)})=-0.13$&$C(\eta_{\rm ew}|V_{cb}|h_{A_1}(1),R_1(1))=-0.108$\\
$C(a_{0}^{(0)},a_{2}^{(0)})=-0.06$&$C(\eta_{\rm ew}|V_{cb}|h_{A_1}(1),R_2(1))=-0.063$\\
$C(a_{1}^{(0)},a_{2}^{(0)})=-0.12$&$C(\rho_{D^*}^2,R_1(1))=0.568$\\
$C(a_{0}^{(0)},V_1(1))=0.50$&$C(\rho_{D^*}^2,R_2(1))=-0.809$\\
$C(a_{1}^{(0)},V_1(1))=0.05$&$C(R_1(1),R_2(1))=-0.758$\\
$C(a_{2}^{(0)},V_1(1))=0.07$&\\
\hline
\end{tabular}
\end{table}

As for the $B D^*$ process, the axial, vector and pseudoscalar form factors are described in terms of a HQET form factor, $h_{A_1}(w)$,
and the ratios $R_i(w)$~\cite{Caprini:1997mu,Fajfer:2012vx}:
\begin{align}
&V(w)=\frac{R_1(w)}{r_{D^*}}h_{A_1}(w)&A_0(w)=\frac{R_0(w)}{r_{D^*}}h_{A_1}(w) &\nonumber\\
&A_1(w)=\frac{w+1}{2}r_{D^*}h_{A_1}(w)&A_2(w)=\frac{R_2(w)}{r_{D^*}}h_{A_1}(w),&\label{eq:ffparVAs}
\end{align}
with 
\begin{align}
h_{A_1}(w)=&h_{A_1}(1)\left[1-8\rho_{D^*}^2z+(53\rho_{D^*}^2-15)z^2-(231\rho_{D^*}^2-91)z^3\right], \nonumber\\
&R_1(w)=R_1(1)-0.12(w-1)+0.05(w-1)^2,\nonumber\\
&R_2(w)=R_2(1)+0.11(w-1)-0.06(w-1)^2,\nonumber\\
&R_0(w)=R_0(1)-0.11(w-1)+0.01(w-1)^2.\label{eq:ffparRs}
\end{align}
The parameters $\rho_{D^*}^2$ and $R_{1,2}(1)$ are obtained from fits to the $\bar B\to D^*\ell\bar\nu$ spectra~\cite{Amhis:2014hma}, $h_{A_1}(1)$ 
can be obtained from lattice calculations~\cite{Bailey:2014tva} and $R_0(1)$ can be calculated using HQET~\cite{Fajfer:2012vx}. 
Finally, the tensor form factors can be related to $h_{A_1}(w)$ at leading order in the heavy-quark and perturbative expansions:
\begin{equation}
T_0(w)=\mathcal{O}(\Lambda/m_Q),\hspace{0.3cm}T_1(w)=\sqrt{m_{D^*}/m_B}\,h_{A_1}(w)+\mathcal{O}(\Lambda/m_Q),
\hspace{0.3cm}T_2(w)=\sqrt{m_{B}/m_{D^*}}\,h_{A_1}(w)+\mathcal{O}(\Lambda/m_Q). \label{eq:ffparTs}
\end{equation}

In Tab.~\ref{tab:FFparam} we list the numerical values for the form factor parameters and total normalizations of the amplitudes
that are employed in this work. For the tensor form factors in eqs.~(\ref{eq:ffNPDTens},~\ref{eq:ffNPDsTens}), we neglect 
the $\Lambda/m_Q$ power corrections so that the sensitivity to the tensor operator in our analyses has a $\sim25\%$ relative uncertainty.
Note that combining this with the values of the HQET form factors at $w=1$ calculated 
in LQCD one obtains $|V_{cb}|=41.2(1.4)_{\rm exp}(1.6)_{\rm th}\times10^{-3}$ and $|V_{cb}|=39.5(0.5)_{\rm exp}(0.6)_{\rm th}\times10^{-3}$ 
for $\bar B\to D\ell^-\bar \nu$ and $\bar B\to D^{*}\ell^-\bar \nu$ respectively.

\subsection{The helicity amplitudes and decay rates into polarized $\tau$}

Neglecting electromagnetic radiative corrections the $B\to D^{(*)}\tau\bar\nu$ amplitude factorizes as:
\begin{eqnarray}
 \mathcal{M}=-\sqrt{2}G_F V_{cb}\left\{H_{V}^\mu\langle\tau^-\bar\nu|\bar\tau\gamma_\mu P_L\nu|0\rangle+H_{S}\langle\tau^-\bar\nu|\bar\tau P_L\nu|0\rangle
 +H_T^{\mu\nu}\langle\tau^-\bar\nu|\bar\tau\sigma_{\mu\nu} P_L\nu|0\rangle\right\},\label{eq:ampBDtaunu}
\end{eqnarray}
with $H_V^\mu$, $H_S$ and $H_T^{\mu\nu}$: 
\begin{align}
&H_V^\mu=\left(1+\epsilon_L^\tau+\epsilon_R\right)\langle \bar{c}\gamma^\mu b\rangle+
\left(\epsilon_R-\epsilon^\tau_L-1\right)\langle \bar{c}\gamma^\mu\gamma_5 b\rangle,\nonumber\\
&H_S=(\epsilon_{S_R}^\tau+\epsilon_{S_L}^\tau)\langle \bar c b\rangle+(\epsilon_{S_R}^\tau-\epsilon_{S_L}^\tau)\langle \bar c\gamma_5 b\rangle,\nonumber\\
&H_T^{\mu\nu}=\epsilon_T^\tau\langle\bar c\sigma^{\mu\nu}(1-\gamma_5)b\rangle,\label{eq:Hampsdef}
\end{align}
subsuming Wilson coefficients and the hadronic matrix elements (schematically denoted as $\langle\ldots\rangle$) of
the different quark bilinears stemming from the effective Lagrangian in eq.~(\ref{eq:leff1}). The contributions to the 
amplitude eq.~(\ref{eq:ampBDtaunu}) can be projected into the different angular-momentum states of the dilepton pair, 
characterized by its polarization vectors $\eta^\mu(\lambda)$ and using the completeness relation 
$g_{\mu\nu}=\sum g_{mn}\eta_\mu(m)\eta_\nu^*(n)$. The projections of $H_V^\mu$ and $H_T^{\mu\nu}$
define the helicity amplitudes,
\begin{equation}
H_\lambda=H_V^{\mu}\eta^*_\mu(\lambda),\hspace{0.5cm}H_{\lambda\lambda^\prime}=H_T^{\mu\nu}\eta^{*}_\mu(\lambda)\eta^{*}_\nu(\lambda^\prime).  \label{eq:helicityDef}
\end{equation}
while $H_S$ contributes only to the $\lambda=t$ component. The tensor helicity amplitudes are antisymmetric with respect to the
exchange of indices, $H_{\lambda^\prime\lambda}=-H_{\lambda\lambda^\prime}$.

\begin{figure}[h]
\begin{tabular}{cc}
  \includegraphics[width=8cm]{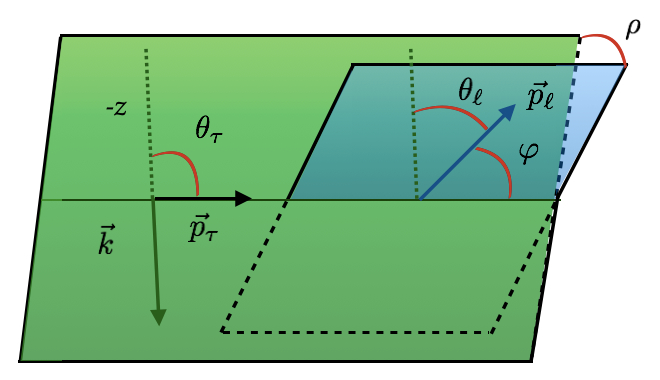}
\end{tabular}
\caption{Kinematics of the chain decay $\bar B\to D^{(*)}\tau^-(\to\ell^-\bar\nu_\ell\nu_\tau)\bar\nu_\tau$.
\label{fig:kinematics}}
\end{figure}

We set the $xyz$ coordinate system  so that $\hat z=\vec k/|\vec k|$ in the $B$- or $q$-rest
frames and with the 3-momentum of the $\tau$, $\vec p_\tau$ contained in the $zx$ plane (see Fig.~\ref{fig:kinematics}). The kinematics of the $B\to D^{(*)}\tau\bar\nu$ decay can then
be fully characterized by $q^2$ and the angle $\theta_\tau$ of $\vec p_\tau$ relative to $-\hat z$ defined in the $q$-rest frame.
The momentum and energy of the $D^{(*)}$ are functions of $q^2$; in the $B$-rest frame:
\begin{align}
E_{D^{(*)}}=\frac{1}{2m_B}(m_B^2+m_{D^{(*)}}^2-q^2),\hspace{0.3cm} q^0=\frac{1}{2m_B}(m_B^2+q^2-m_{D^{(*)}}^2),\hspace{0.3cm}
|\vec{k}|=\frac{1}{2m_B}\sqrt{\lambda(m_B^2,m_{D^{(*)}}^2,q^2)}.
\end{align}
where $\lambda(x,y,z)=a^2+b^2+c^2-2(ab+ac+bc)$. For the dilepton polarization vectors we use 
$\eta^\mu(\pm)=(0,\pm1,-i,0)/\sqrt{2}$, $\eta^\mu(0)=(|\vec k|,0,0,-q^0)/\sqrt{q^2}$ and 
$\eta^\mu(t)=(q^0,0,0,-|\vec k|)/\sqrt{q^2}$ and for the $D^*$, $\epsilon^\mu(\pm)=(0,\mp1,-i,0)/\sqrt{2}$ and
$\epsilon^\mu(0)=(|\vec{k}|,0,0,E_{D^*})/m_{D^*}$.

Conservation of angular momentum implies that the only non-vanishing
helicity amplitudes for the $B\to D\tau\bar\nu$ decay are those which project into $\lambda=0,t$ (or $\lambda+\lambda^\prime=0$ for the tensors): 
\begin{align}
&H_0=(1+\epsilon_L^\tau+\epsilon_R)\frac{2m_B|\vec k|}{\sqrt{q^2}}f_+(q^2),\hspace{0.2cm}H_t=(1+\epsilon_L^\tau+\epsilon_R)\frac{m_B^2-m_D^2}{\sqrt{q^2}}f_0(q^2),\nonumber\\
&H_S=(\epsilon_{S_R}^\tau+\epsilon_{S_L}^\tau)\frac{m_B^2-m_D^2}{m_b-m_c}f_0(q^2),\hspace{0.3cm}H_{+-}=-H_{t0}=\epsilon_T^\tau\frac{2im_B|\vec k|}{m_B+m_D} f_T(q^2).\label{eq:helicityBD}
\end{align}
For the decay into polarized $D^*(\tilde\lambda)$, conservation of angular momentum requires 
that the only components that contribute to the amplitude are $\lambda=\pm$ for $\tilde\lambda=\pm$ and $\lambda=0,\,t$ for $\tilde\lambda=0$ or, 
in case of the projections of $H_T^{\mu\nu}$, $\lambda+\lambda^\prime=\tilde\lambda$. The helicity amplitudes
evaluated in the $B$-rest frame are:
\begin{align}
&H_\pm=-(1+\epsilon^\tau_L-\epsilon_R)(m_B+m_{D^*})A_1(q^2)\pm(1+\epsilon_L^\tau+\epsilon_R)\frac{2m_B|\vec k|}{m_B+m_{D^*}}V(q^2),\nonumber\\
&H_0=-\frac{1+\epsilon_L^\tau-\epsilon_R}{2 m_{D^*}\sqrt{q^2}}\left[\left(m_B+m_{D^*}\right)\left(m_B^2-m_{D^*}^2-q^2\right)A_1(q^2)
-\frac{4m_B^2|\vec k|^2}{m_B+m_{D^*}}A_2(q^2)\right],\nonumber\\
&H_t=-(1+\epsilon_L^\tau-\epsilon_R)\frac{2m_B|\vec k|}{\sqrt{q^2}}A_0(q^2),\hspace{0.5cm}H_S=(\epsilon_{S_R}-\epsilon_{S_L})\frac{2m_B|\vec k|}{m_b+m_c}A_0(q^2),\nonumber\\
&H_{\pm0}=\pm\frac{i\epsilon_T^\tau}{2\sqrt{q^2}}\left[(m_B^2-m_{D^*}^2\pm2m_B|\vec k|)(T_1(q^2)+T_2(q^2))+q^2(T_1(q^2)-T_2(q^2))\right],\nonumber\\
&H_{\pm t}=\frac{i\epsilon_T^\tau}{2\sqrt{q^2}}\left[(m_B^2-m_{D^*}^2\pm2m_B|\vec k|)(T_1(q^2)+T_2(q^2))+q^2(T_1(q^2)-T_2(q^2))\right],\nonumber\\
&H_{+-}=-H_{t0}=i\epsilon_T^\tau\left[\frac{m_BE_{D^*}}{m_{D^*}}T_1(q^2)+m_{D^*}T_2(q^2)+\frac{m_B^2|\vec k|^2}{m_{D^*}(m_{D^*}+m_B)^2}T_0(q^2)\right].\label{eq:helicityBDV}
\end{align}

The differential decay rate for $B\to D^{(*)}\tau\bar\nu$ for a $\tau$ polarized along a particular direction
$\hat s$ is~\cite{Tanaka:1994ay}:
\begin{align}
d\Gamma_B(\hat s)=\frac{1}{2}\left[d\Gamma_B+\left(d\Gamma_B^L\,\hat z^\prime+d\Gamma_B^\perp\,\hat x^\prime+d\Gamma_B^T\,\hat y^\prime\right)\cdot \hat s\right], \label{eq:Gammaspoln}
\end{align}
where we have introduced a second coordinate system denoted by $x^\prime y^\prime z^\prime$ set in the $q$ rest-frame and defined by:
\begin{align}
\hat z^\prime=\frac{\vec p_\tau}{|\vec p_\tau|}=\sin\theta_\tau\,\hat x-\cos\theta_\tau\,\hat z,\hspace{0.5cm}\hat y^\prime=\hat y=\frac{\vec k\times\vec p_\tau }{|\vec k||\vec p_\tau|}, \hspace{0.5cm}
\hat x^\prime=\hat y^\prime\times \hat z^\prime=-\cos\theta_\tau\,\hat x-\sin\theta_\tau\,\hat z. \label{eq:basistrans}
\end{align}
The different contributions to the decay rate in eq.~(\ref{eq:Gammaspoln}) are, on one hand:
\begin{align}
d\Gamma_{B}= d\Gamma_{B,+}+d\Gamma_{B,-},\hspace{0.5cm}d\Gamma_{B}^L= d\Gamma_{B,+}-d\Gamma_{B,-},
\end{align}
with the $d\Gamma_{B,\pm}$ the differential decay rates corresponding to the two helicities  of the $\tau$, $\lambda_\tau=\pm1/2$. On the other hand,
for the components orthogonal to $\vec p_\tau$ we have interference effects:
\begin{align}
 d\Gamma_B^\perp=\frac{(2\pi)^4\,d \Phi_3}{2 m_B}2{\rm Re}\left[\mathcal M_{B+}\mathcal M_{B-}^\dagger\right],\hspace{0.5cm}
 d\Gamma_B^T=\frac{(2\pi)^4\,d \Phi_3}{2 m_B}2{\rm Im}\left[\mathcal M_{B+}\mathcal M_{B-}^\dagger\right],
\end{align}
where $\mathcal M_{B\pm}$ is the amplitude of the $B\to D^{(*)}\tau\bar\nu$ decay for $\lambda_\tau=\pm1/2$ and 
$d \Phi_3\equiv d \Phi_3(p;k,p_{\bar\nu_\tau},p_\tau)$ is the corresponding 3-body phase space differential element.

Solving the phase space in terms of the kinematic variables introduced above (Fig.~\ref{fig:kinematics}) for the rates involving a
given helicity of the $\tau$ leads to:
\begin{align}
\frac{d^2\Gamma_{B,+}}{dq^2d(\cos\theta_\tau)}&=\frac{G_F^2|V_{cb}|^2\eta_{\rm ew}^2}{256\pi^3}\frac{|\vec{k}|}{m_B^2}\left(1-\frac{m_\tau^2}{q^2}\right)^2
\left\{2\cos^2\theta_\tau\,\Gamma_{0+}^0+2\Gamma_{0+}^t+2\cos\theta_\tau\,\Gamma_{0+}^I+
\sin^2\theta_\tau(\Gamma_{++}+\Gamma_{-+})\right\},\nonumber \\
\frac{d^2\Gamma_{B,-}}{dq^2d(\cos\theta_\tau)}&=\frac{G_F^2|V_{cb}|^2\eta_{\rm ew}^2}{256\pi^3}\frac{|\vec{k}|}{m_B^2}\left(1-\frac{m_\tau^2}{q^2}\right)^2
\left\{2\sin^2\theta_\tau\Gamma_{0-}+(1-\cos\theta_\tau)^2\Gamma_{+-}+(1+\cos\theta_\tau)^2\Gamma_{--}\right\}.\label{eq:BDtaunurate}
\end{align}
where we have separated the contributions coming from the different $D^{(*)}$ and $\tau$ helicity states, 
$\Gamma_{\tilde \lambda,\lambda_\tau}$. For the decay into a $\tau(1/2)$ and a longitudinal $D^{(*)}$ we separate the three
contributions stemming from the longitudinal ($\Gamma_{0+}^0$) and time-like ($\Gamma_{0+}^t$) components of the dilepton state and from
their interference ($\Gamma_{0+}^I$). Note that for the $BD$ channel all the contributions from the transversal components are equal to 0. 
Finally, the $\Gamma^{(X)}_{\tilde \lambda,\lambda_\tau}$ are functions of the helicity amplitudes:
\begin{align}
\Gamma_{0+}^0&=\left|2i\sqrt{q^2}\left(H_{+-}+H_{0t}\right)-m_\tau H_0\right|^2,\hspace{0.5cm}\Gamma_{0+}^t=\left|m_\tau H_t+\sqrt{q^2}H_S\right|^2,\nonumber\\
&\Gamma_{0+}^I=2{\rm Re}\left[(2i\sqrt{q^2}\left(H_{+-}+H_{0t}\right)-m_\tau H_0)(m_\tau H_t+\sqrt{q^2}H_S)^*\right],\nonumber\\
\Gamma_{++}&=\left|m_\tau H_+-2i\sqrt{q^2}\left(H_{+t}+H_{+0}\right)\right|^2,\hspace{0.5cm}\Gamma_{-+}=\left|m_\tau H_--2i\sqrt{q^2}\left(H_{-t}-H_{-0}\right)\right|^2,\nonumber\\
&\hspace{2cm}\Gamma_{0-}=\left|\sqrt{q^2}H_0-2im_\tau\left(H_{+-}+H_{0t}\right)\right|^2,\nonumber\\
\Gamma_{+-}&=\left|\sqrt{q^2} H_+-2im_\tau\left(H_{+t}+H_{+0}\right)\right|^2,\hspace{0.2cm}
\Gamma_{--}=\left|\sqrt{q^2} H_--2im_\tau\left(H_{-t}-H_{-0}\right)\right|^2. \label{eq:Gammashel}
\end{align}

Likewise, the contribution to the rate of the interference term $d\mathcal I_B\equiv(2\pi)^4\,d \Phi_3/(2 m_B)\mathcal M_{B+}\mathcal M_{B-}^\dagger$ is:
\begin{align}
\frac{d^2\mathcal I_B}{dq^2d(\cos\theta_\tau)}&=\frac{G_F^2|V_{cb}|^2\eta_{\rm ew}^2}{256\pi^3}\frac{|\vec{k}|}{m_B^2}\left(1-\frac{m_\tau^2}{q^2}\right)^2\sin\theta_\tau
\left[2\,\mathcal I_0\cos\theta_\tau+2\,\mathcal I_{0}^I+\mathcal I_+(1-\cos\theta_\tau)+\mathcal I_-(1+\cos\theta_\tau)\right],\label{eq:GammasInt}
\end{align}
where
\begin{align}
\mathcal I_0&=m_\tau\sqrt{q^2}|H_0|^2+4m_\tau\sqrt{q^2}|H_{+-}+H_{0t}|^2+2i\,m_\tau^2H_0(H_{+-}+H_{0t})^*
-2i\,q^2H_0^*(H_{+-}+H_{0t}),\nonumber\\
\mathcal I^I_0&=-\sqrt{q^2}H_0^*(m_\tau H_t+\sqrt{q^2}H_S)-2i\,m_\tau (H_{+-}+H_{0t})^*
(m_\tau H_t+\sqrt{q^2}H_S),\nonumber\\
\mathcal I_+&=m_\tau\sqrt{q^2}|H_+|^2+2i\,m_\tau^2H_+(H_{+t}+H_{+0})^*-2i\,q^2H_+^*(H_{+t}+H_{+0})+4m_\tau\sqrt{q^2}|H_{+t}+H_{+0}|^2,\nonumber\\
\mathcal I_-&=-m_\tau\sqrt{q^2}|H_-|^2-2i\,m_\tau^2H_-(H_{-t}-H_{-0})^*+2i\,q^2H_-^*(H_{-t}-H_{-0})-4m_\tau\sqrt{q^2}|H_{-t}-H_{-0}|^2.\nonumber\\
\end{align}
The differential forms for the $\Gamma_{B}^{\perp,T}$ observables are obtained taking twice the real or imaginary part in eq.~(\ref{eq:GammasInt}). Note 
that the latter observable is a triple-product correlation that, in the absence of final-state interactions, is $T$-odd and becomes sensitive to NP sources 
of $CP$-violation entering in the process through the Wilson coefficients~\cite{Tanaka:1994ay}.

\section{The $\bar B\to D^{(*)}\tau^-(\to\ell^-\bar\nu_\ell\nu_\tau)\bar\nu_\tau$ decay rate}
\label{sec:5body}
\subsection{The leptonic $\tau$ decay}

In the SM, the differential decay rate $\tau^-\to\ell^-\bar\nu_\ell\nu_\tau$ with the $\tau$ lepton with helicity $\lambda_\tau$ is: 
\begin{align}
&d\Gamma_{\tau,\lambda_\tau}=\frac{32G_F^2(2\pi)^4}{m_\tau}(p_{\nu_\tau}\cdot p_\ell)\left[p_{\bar\nu_\ell}\cdot(p_\tau-m_\tau s_{\lambda_\tau})\right]
\,\delta^{(4)}(p_\tau-p_\ell-p_{\bar \nu_\ell}-p_{\nu_\tau})
\frac{d^3\vec p_\ell}{2E_\ell(2\pi)^3}\frac{d^3\vec p_{\bar\nu_\ell}}{2E_{\bar\nu_\ell}(2\pi)^3}\frac{d^3\vec p_{\nu_\tau}}{2E_{\nu_\tau}(2\pi)^3},\label{eq:taudecay0}
\end{align}
where $s_{\lambda_\tau}$ is the spin 4-vector of the $\tau$. Integrating over the phase space of the neutrinos,
\begin{equation}
\int \frac{d^3\vec p_{\bar\nu_\ell}}{2E_{\bar\nu_\ell}(2\pi)^3}\frac{d^3\vec p_{\nu_\tau}}{2E_{\nu_\tau}(2\pi)^3}p_{\bar\nu_\ell}^\alpha p_{\nu_\tau}^\beta
\delta^{(4)}(p_\tau-p_\ell-p_{\bar \nu_\ell}-p_{\nu_\tau})=\frac{1}{48(2\pi)^5}\left[(p_{\tau}-p_\ell)^2g^{\alpha\beta}
+2(p_{\tau}-p_\ell)^\alpha(p_{\tau}-p_\ell)^\beta\right], \label{eq:neutrinosPSI}
\end{equation}
one obtains,
\begin{equation}
d\Gamma_{\tau,\lambda_\tau}=\frac{G_F^2}{3(2\pi)^4}\frac{d^3\vec p_\ell}{m_\tau E_\ell}p_{\ell,\alpha}\left(p_\tau-m_\tau s_{\lambda_\tau}\right)_\beta
\left[(p_{\tau}-p_\ell)^2g^{\alpha\beta}+2(p_{\tau}-p_\ell)^\alpha(p_{\tau}-p_\ell)^\beta\right].\label{eq:taudecayLorentz}
\end{equation}
For the sake of simplicity, we will assume in the following that $m_\ell=0$, which should hold with good accuracy
for most of the kinematics of the decays considered in this work. 

To resolve the $\tau$ decay in $\bar B\to D^{(*)}\tau^-(\to\ell^-\bar\nu_\ell\nu_\tau)\bar\nu_\tau$, it is convenient
to use the coordinate system $x^\prime y^\prime z^\prime$ introduced above.
The spherical coordinates of $\vec p_\ell$ in this basis are the polar angle $\varphi$ with respect to $\hat z^\prime$, and the
corresponding azimuthal angle $\rho$ with respect to the $x^\prime z^\prime$ plane (see Fig.~\ref{fig:kinematics}). In this coordinate system,
using eq.~(\ref{eq:taudecayLorentz}) in the $q$ rest-frame, we obtain:
\begin{align}
\frac{d\Gamma_{\tau,\pm}}{dE_\ell d(\cos\varphi)d\rho}=&\frac{G_F^2}{3(2\pi)^4}\frac{E_\ell^2}{m_\tau}
\left[(E_\tau-|\vec p_\tau|\cos\varphi)(3m_\tau^2-4 E_\ell (E_\tau-|\vec p_\tau|\cos\varphi)) \right.\nonumber\\
&\left.\pm(E_\tau \cos\varphi-|\vec p_\tau|)(m_\tau^2-4 E_\ell( E_\tau-
|\vec p_\tau| \cos\varphi)\right],\label{eq:taudecayboosted}
\end{align}
where $E_\tau=(q^2+m_\tau^2)/(2\sqrt{q^2})$ and $|\vec p_\tau|=(q^2-m_\tau^2)/(2\sqrt{q^2})$. Integrating in all the phase space and averaging over polarizations
we find that the branching fraction of the $\tau^-\to\ell^-\nu_\tau\bar\nu_\ell$ decay is:
\begin{equation}
\mathcal{B}[\tau_\ell]= \tau_\tau \frac{G_F^2m_\tau^5}{192\pi^3},\label{eq:tauBR}
\end{equation}
where $\tau_\tau$ is the $\tau$-lepton lifetime $\tau_\tau=1/\Gamma_\tau$. This expression leads, numerically,
to $\mathcal{B}[\tau_\ell]= 0.178$ which, at the level of precision of this study, is well in agreement with the experimental data~\cite{Agashe:2014kda}.

Like in the case of the $B$ decay, one may also study the decays rates for $\tau$ with the spin pointing to an arbitrary direction. This
will involve, in general, interference effects between the $\tau$ helicity decay amplitudes, $\mathcal M_{\tau\pm}$. Defining the contribution
of these terms to the rate as $d\mathcal I_\tau\equiv(2\pi)^4\,d \Phi_3(p_\tau;\,p_\ell,\,p_{\bar \nu_\ell})/(2 m_B)\mathcal M_{\tau+}\mathcal M_{\tau-}^\dagger$, 
one obtains:
\begin{align}
\frac{ d^3\mathcal I_\tau}{dE_\ell d(\cos\varphi)d\rho}=\frac{G_F^2}{3(2\pi)^4}E_\ell^2\left[e^{i\,\rho}\sin\varphi\left(m_\tau^2-4E_\ell(E_\tau-|\vec p_\tau|\cos\varphi)\right)\right]. \label{eq:tauinterf}
\end{align}

\subsection{The 5-body differential decay rate}

The $\bar B\to D^{(*)}\tau^-(\to\ell^-\bar\nu_\ell\nu_\tau)\bar\nu_\tau$ decay amplitude is: 
\begin{align}
 \mathcal{M}=&\frac{4G_F^{2} V_{cb}\eta_{\rm ew}}{p_\tau^2-m_\tau^2+i\,m_\tau\Gamma_\tau}\sum_{\lambda_\tau=\pm1/2}\langle\ell^-\bar\nu_\ell|\bar\ell\gamma^\rho P_L\nu_\ell|0\rangle
 \langle\nu_\tau|\bar\nu_\tau\gamma_\rho P_L\tau|\tau^-(\lambda_\tau)\rangle\times\nonumber\\
 &\times\left\{H_{V}^\mu\langle\tau^-(\lambda_\tau)\bar\nu_\tau|\bar\tau\gamma_\mu P_L\nu_\tau|0\rangle+H_{S}\langle\tau^-(\lambda_\tau)\bar\nu_\tau|\bar\tau P_L\nu_\tau|0\rangle
 +H_T^{\mu\nu}\langle\tau^-(\lambda_\tau)\bar\nu_\tau|\bar\tau\sigma_{\mu\nu} P_L\nu_\tau|0\rangle\right\},\label{eq:ampBDellnununu}
\end{align}
where we have used the completeness relation $\slashp_\tau+m_\tau=\sum_{\lambda_\tau}\,u(p_\tau,\lambda_\tau)\bar u(p_\tau,\lambda_\tau)$ and the 
amplitude factorizes into the $\bar B\to D^{(*)}\tau^-(\lambda_\tau)\bar\nu_\tau$ and $\tau^-(\lambda_\tau)\to\ell^-\bar\nu_\ell\nu_\tau$ amplitudes. 
The phase space differential volume of the 5-body decay also factorizes into those of the two 3-body decays according to the formula:
\begin{equation}
d\Phi_5(p;\,k,\,p_{\bar \nu_\tau},\,p_\ell,\,p_{\bar \nu_\ell},\,p_{\nu_\tau})=(2\pi)^3 d\Phi_3(p;\,k,\,p_{\bar \nu_\tau},\,p_\tau)
d\Phi_3 (p_\tau;\,p_\ell,\,p_{\bar \nu_\ell},\,p_{\nu_\tau}) dp_\tau^2. \label{eq:psfactor}
\end{equation}
In the narrow-width approximation, $\Gamma_\tau\ll m_\tau$, which applies to an excellent degree here:
\begin{align}
\frac{1}{(p_\tau^2-m_\tau^2)^2+m_\tau^2\Gamma_\tau^2}\xrightarrow{\Gamma_\tau\ll m_\tau}\frac{\pi}{m_\tau\Gamma_\tau}\delta(p_\tau^2-m_\tau^2),\label{eq:nwapp}
\end{align}
the $\tau$ is on-shell, resolving the phase-space integral in $dp_\tau^2$. Combining eqs.~(\ref{eq:ampBDellnununu},~\ref{eq:psfactor},~\ref{eq:nwapp})
we arrive at:
\begin{align}
d\Gamma=&\tau_\tau\,\sum_{\lambda_\tau=\pm1/2}d\Gamma_{B,\lambda_\tau}\times d\Gamma_{\tau,\lambda_\tau}+\tau_\tau d\mathcal I_B\times d\mathcal I_\tau
+{\rm c.c.}\\
=&\tau_\tau\,\sum_{\lambda_\tau=\pm1/2}d\Gamma_{B,\lambda_\tau}\times d\Gamma_{\tau,\lambda_\tau}+
\tau_\tau\left(\cos\rho\,d\Gamma_B^\perp-\sin\rho\,d\Gamma_B^T\right)d|\mathcal I_\tau|.\label{eq:factrates}
\end{align} 
where, in the second line, we have used eq.~(\ref{eq:tauinterf}) and introduced the polarization observables in eq.~(\ref{eq:Gammaspoln}) .
The 5-body differential decay rate $d\Gamma$ is a function of $q^2$, $E_\ell$ and the angular variables 
$\rho$, $\cos\theta_\ell$ and $\cos\varphi$ introduced above and shown in Fig.~\ref{fig:kinematics}. 
It is relevant to point out that $\rho$ is the only variable in the expression above linking the production and decay systems. In particular
it introduces a correlation between the production $D^{(*)}-\tau$ plane and the decay
$\tau-\ell$ plane. One can also see from Eq.~(\ref{eq:factrates}) that after integration in $\rho\in(-\pi,\pi)$ the interference
term vanishes and the intuitive implementation of the narrow width approximation holds~\cite{Dicus:1984fu}. Nevertheless, since the present work will
discuss angular distributions the interference term does have an effect.

\subsubsection{Integrating the $\tau$ angular phase-space}

Experiments can, at best, measure the distribution of decays with respect to the variables $q^2$, $E_\ell$ and the
angle of the 3-momentum of this final-state lepton relative to the one of the $D^{(*)}$, that we define analogously to $\theta_\tau$, in the $q$ rest frame, as:
\begin{align}
\cos\theta_\ell=-\frac{\vec p_\ell\cdot \vec{k}}{|\vec p_\ell||\vec k|}=-\frac{\vec p_\ell\cdot \hat z}{|\vec p_\ell|}=
\cos\theta_\tau\cos\varphi+\sin\theta_\tau\sin\varphi\cos\rho, \label{eq:defcosthetal}
\end{align}
We now need to integrate the expression of the 5-body differential decay rate in eq.~(\ref{eq:factrates}) only 
in the angular phase space of the $\tau$. To do this, we use eq.~(\ref{eq:defcosthetal}) to transform the angular
variables:
\begin{align}
(\rho,\,\cos\theta_\tau,\,\cos\varphi)\to(\cos\theta_\ell,\,\cos\theta_\tau,\,\cos\varphi), \label{eq:transfangvar}
\end{align}
and we integrate in $\cos\theta_\tau$ and in $\cos\varphi$ for a given $\cos\theta_\ell$.
Before we proceed, note that this transformation maps the domain of integration $\Theta$ defined by $\rho\in[-\pi,\,0]\cup[0,\,\pi]$ 
and $\cos\theta_\tau\in[-1,\,1]$ twice onto $\Theta^\prime$ delimited by $\cos\theta_\ell\in[-1,\,1]$ and 
$\cos\theta^\pm_\tau=\cos(\theta_\ell\mp\varphi)$. The contribution to the decay rate from any differential of phase space 
in $\Theta^\prime$  is then related to the sum of the corresponding ones in $\Theta$, which are themselves related by 
$d\rho\,f(\rho)|_{[-\pi,\,0]}=d\rho\,f(-\rho)|_{[0,\,\pi]}$ and where, as shown in eq.~(\ref{eq:factrates}), $f(\rho)$ can be $1$, $\cos\rho$
or $\sin\rho$. Therefore, the contributions from decay rates without interference $d\Gamma_{B,\lambda_\tau}\times d\Gamma_{\tau,\lambda_\tau}$ 
and from $d\Gamma_B^\perp$ should be multiplied by a factor 2 when integrating over $\Theta^\prime$. On the other hand, the \textit{contribution 
of the $CP$-odd observable $d\Gamma_B^T$ vanishes from the angular distribution in $\cos\theta_\ell$.} This can be understood noticing that the
relative $D^{(*)}-\tau$ angle is the same for $\rho$ and $-\rho$, whereas the CP odd contribution changes sign.

\begin{table}[h]
\centering
\caption{Results for all the nonvanishing angular integrals $I_f(\theta_\ell,\,\varphi)$ defined in eq.~(\ref{eq:integthetatauniv}). 
\label{tab:integthetauniv}}
\begin{tabular}{|c|c|c|c|c|c|}
\hline
$f(\theta_\tau)$&$\;\;\;1\;\;\;$&$\cos\theta_\tau$&$\cos^2\theta_\tau$&$\cos\rho\,\sin\theta_\tau$&$\cos\rho\,\sin(2\theta_\tau)$\\
\hline
$I_f(\theta_\ell,\varphi)$&$\pi$&$\pi\cos\theta_\ell\cos\varphi$&$\pi\left(\cos^2\theta_\ell\cos^2\varphi+\frac12\sin^2\theta_\ell\sin^2\varphi\right)$&
$\pi\cos\theta_\ell\sin\varphi$&$\pi\sin\varphi\cos\varphi(3\cos^2\theta_\ell-1)$\\
\hline
\end{tabular}
\end{table} 

Now we turn to the integration on $\theta_\tau$. These integrals are of the form:
\begin{align}
I_f(\theta_\ell,\,\varphi)=\int_{\cos\theta^-_\tau}^{\cos\theta^+_\tau}d(\cos\theta_\tau)\left|\det {\bf J}\right|\,f(\theta_\tau),\label{eq:integthetatauniv}
\end{align}
where $f(\theta_\tau)$ is a given function and $\det \bf{J}$ is the determinant of the Jacobian
of the transformation in eq.~(\ref{eq:transfangvar}):
\begin{align}
\det {\bf J}=-\frac{1}{\sin\rho\sin\theta_\tau\sin\varphi}=-(1-\cos^2\theta_\tau-\cos^2\varphi-\cos^2\theta_\ell+2\cos\theta_\tau\cos\varphi\cos\theta_\ell)^{-1/2}\,,\label{eq:difftransrho} 
\end{align}
The dependence of the rate in $\theta_\tau$  enters through the Jacobian or through $f(\theta_\tau)$, which encompasses 
the angular dependence of the $B\to D^{(*)}\tau^-\bar\nu_\tau$ rates, eqs.~(\ref{eq:BDtaunurate},~\ref{eq:GammasInt}) 
or of $\cos\rho$ via eq.~(\ref{eq:defcosthetal}) in case of the interference 
term in eq.~(\ref{eq:factrates}).~\footnote{Note that this is one of the benefits of the $q$ rest frame; 
in a different frame $E_\tau$ in eq.~(\ref{eq:taudecayboosted}) depends on $\theta_\tau$.
the results for all the integrals appearing in our case. 
}  In Tab.~\ref{tab:integthetauniv} we collect the results for the different (nonvanishing) integrals that appear. In consistency with
the discussion above regarding the contribution of the interference term $d\Gamma_B^\perp$ to the rate,  
the last two columns vanish when integrated over the full range of $\cos\theta_\ell$. 

The integral in the angular variable $\cos\varphi$ is also subtle. The energy of the final charged-lepton in
the $\tau$ rest frame, $\tilde E_\ell$ and the one in the $q$ rest-frame are related by the boost:
\begin{equation}
\tilde E_\ell=\gamma(E_\ell-\beta \cos\varphi), \label{eq:boostE}
\end{equation}
where $\gamma=E_\tau/m_\tau$ and $\gamma\beta=|\vec p_\tau|/m_\tau$. Thus, this integral involves non-flat boundaries in
the phase-space variables $E_\ell$ and $\varphi$ that account for the fact that some energy configurations in the $q$-rest frame can
only be reached for certain polar angles $\varphi$, e.g. the maximum possible energy for the lepton, $E^{\rm max}_\ell=\sqrt{q^2}/2$, can only be reached when it
is aligned with the $\tau$ momenta so that the relativistic $\gamma$ factor is the largest. More generally, 
there are two regions of integration:
\begin{equation}
E_{\ell}\in\left[\frac{m_\tau^2}{2\sqrt{q^2}},\,\frac{\sqrt{q^2}}{2}\right],
\hspace{0.5cm}\cos\varphi\in[\frac{1}{\beta}-\frac{m_\tau}{2\gamma\beta E_\ell},\,1]. \label{eq:region1v2} 
\end{equation}
where for every $E_\ell$ there is only an angle with respect to the boost direction $\hat z^\prime$, over which we integrate. The second is the region, 
\begin{equation}
 E_{\ell}\in\left[0,\,\frac{m_\tau^2}{2\sqrt{q^2}}\right],\hspace{0.5cm}\cos\varphi\in[-1,1],\label{eq:region2v1}
\end{equation}
that covers the maximum energies that can be reached by all the polar angles $\varphi$.

We can now write the experimentally accessible 3-fold 5-body differential decay rate as:
\begin{align}
\frac{d^3\Gamma_5}{dq^2dE_\ell d(\cos\theta_\ell)}=\mathcal{B}[\tau_\ell]\,\frac{G_F^2|V_{cb}|^2\eta_{\rm ew}^2}{32\pi^3}\frac{|\vec{k}|}{m_B^2}\left(1-\frac{m_\tau^2}{q^2}\right)^2
\frac{E_\ell^2}{m_\tau^3}\times\left[I_0(q^2,\,E_\ell)+I_1(q^2,\,E_\ell)\cos\theta_\ell+I_2(q^2,\,E_\ell)\cos^2\theta_\ell\right],\label{eq:decayratetotal3}
\end{align}
where the different angular coefficients are functions of $q^2$ and $E_\ell$. The angle $\theta_\ell$ is in the interval $[0,\,\pi]$, whereas for $q^2$ and $E_\ell$
we have two distinct regions of phase space corresponding to the two regions of integration above. Given $q^2$ in the interval $[m_\tau^2,\,(m_B-m_{D^{(*)}})^2]$, the decay rate
as a function of $E_\ell$ is defined piecewise over two different domains: one is the region of phase space where $E_{\ell}\in\left[m_\tau^2/(2\sqrt{q^2}),\,\sqrt{q^2}/2\right]$, 
that we call $\omega_1$, and the second one corresponds to $E_{\ell}\in\left[0,\,m_\tau^2/(2\sqrt{q^2})\right]$ or $\omega_2$. We 
plot these regions and the show the resulting expressions for the $I_i(q^2,\,E_\ell)$ in the Appendix~\ref{sec:appendix}.

The angular coefficients in eq.~(\ref{eq:decayratetotal3}) are new observables that are complementary to the total rates. For instance, the angle can be integrated:
\begin{align}
\frac{d^2\Gamma_5}{dq^2dE_\ell}=\mathcal{B}[\tau_\ell]\,\frac{G_F^2|V_{cb}|^2\eta_{\rm ew}^2}{16\pi^3}\frac{|\vec{k}|}{m_B^2}\left(1-\frac{m_\tau^2}{q^2}\right)^2 \frac{E_\ell^2}{m_\tau^3}
\times\left[I_0(q^2,\,E_\ell)+\frac{1}{3}I_2(q^2,\,E_\ell)\right], \label{eq:decayratetotal2}
\end{align}
showing that the total rates only depend on the functions $I_{0,2}(q^2,\,E_\ell)$. This implies that we would obtain a completely independent observable by measuring the coefficient
$I_{1}(q^2,\,E_\ell)$, which could be done defining a forward-backward asymmetry with respect to the angle $\theta_\ell$:
\begin{align}
\frac{d^2A_{FB}(q^2,\,E_\ell)}{dq^2d E_\ell}=\left(\int^1_{0}d(\cos\theta_\ell)-\int^0_{-1}d(\cos\theta_\ell)\right)\frac{d^3\Gamma_5}{dq^2dE_\ell d(\cos\theta_\ell)}.\label{eq:AFB}
\end{align}
An interesting integrated observable that could be constructed using this forward-backward asymmetry is:
\begin{align}
 R^{(*)}_{FB}=\frac{1}{\mathcal{B}[\tau_\ell]}\frac{1}{\Gamma_{\rm norm.}} A_{FB}, \label{eq:RFB}
\end{align}
which is labeled according to whether it corresponds the $BD$ ($R_{FB}$) or $BD^*$ ($R^*_{FB}$) channel. 
In these definitions we have normalized with the total rate of the normalization decay $B\to D^{(*)}\ell\bar\nu_\ell$, $\Gamma_{\rm norm.}$, and 
the branching fraction of the leptonic $\tau$-decay. The third element in eq.~(\ref{eq:RFB}), $A_{FB}$, is the integrated observable in
eq.~(\ref{eq:AFB}) that could be obtained experimentally subtracting the number of total events in the backward and forward directions. 

\section{Phenomenology}

\begin{figure}[h]
\begin{tabular}{cc}
  \includegraphics[width=78mm]{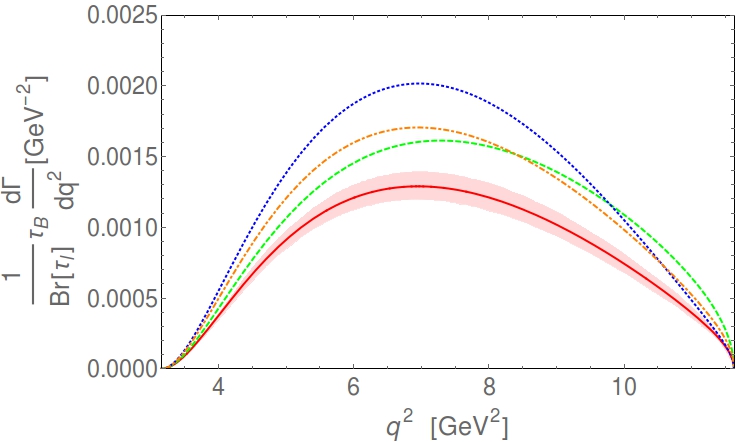}\hspace{0.2cm} &  \includegraphics[width=82mm]{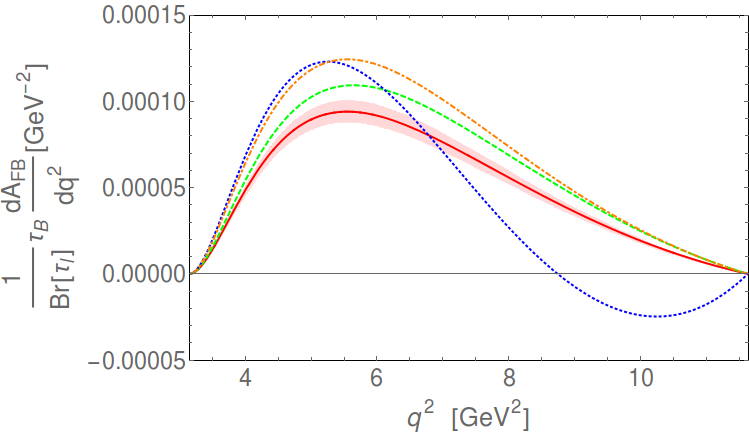} \\
  \includegraphics[width=78mm]{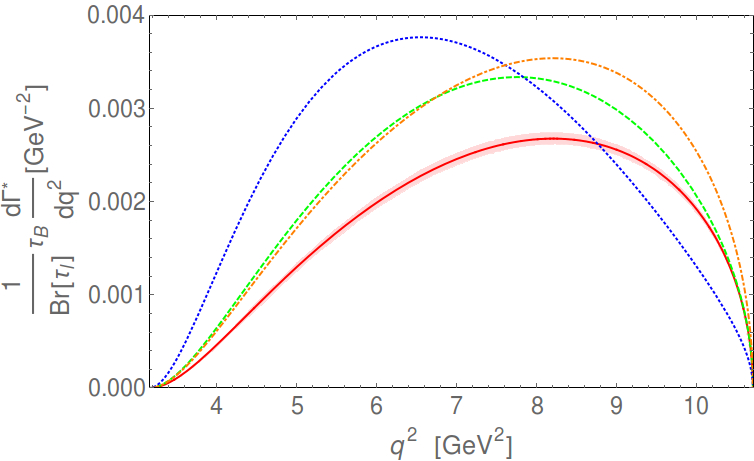}\hspace{0.2cm} &  \includegraphics[width=81mm]{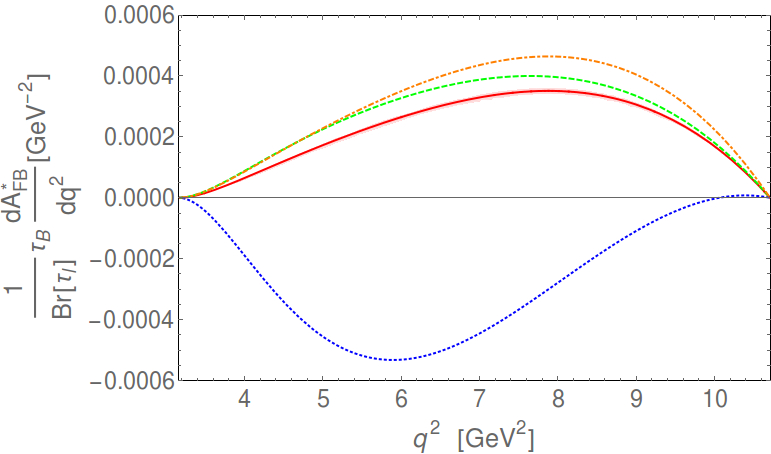}
\end{tabular}
\caption{Decay rates and forward-backward asymmetries, as defined in eq.~(\ref{eq:AFB}), normalized as $B^\pm$ life-time and factoring out
the branching fraction of the leptonic $\tau$ decay. Along with the SM prediction plotted in solid red, we show the results in the three 
benchmark scenarios of NP (see main text): ``Current'' as the dot-dashed (orange) curve, ``Scalar'' as the dashed (green) curve and ``Tensor'' as
the dotted (blue) one. The uncertainties of the SM predictions correspond to the 1$\sigma$ intervals assuming (correlated) Gaussian distributions for the inputs listed in
Tab.~\ref{tab:FFparam} 
\label{fig:diffdecays}}
\end{figure}

In Fig.~\ref{fig:diffdecays} we show the $q^2$-spectrum of the the total rates and the forward-backward asymmetries normalized as 
branching-fractions and where we have factored out the $\mathcal{B}[\tau_\ell]$. The observables of 
the $BD$ and $BD^*$ modes are labeled as the $\Gamma$ ($A_{FB}$) and $\Gamma^*$ ($A^*_{FB}$), respectively. The uncertainties in the SM
predictions correspond to the 1$\sigma$ intervals assuming (correlated) Gaussian distributions for the inputs listed in
Tab.~\ref{tab:FFparam}. Along with these predictions we show the results in three  different
benchmark scenarios of NP. In the first NP scenario, denoted as ``Current'' we only consider a modification of the normalization of the decay via
$\epsilon_L^\tau=0.15$. In the ``Scalar'' scenario we set all $\epsilon_i=0$ except for $\epsilon^\tau_{S_L}=0.80$ and $\epsilon^\tau_{S_R}=-0.65$, while 
in the ``Tensor'' one we only allow for $\epsilon^\tau_{T}=0.40$. Scenarios of this type could explain the $R_{D^{(*)}}$
anomalies with NP at a scale of $\Lambda\sim1$ TeV, as discussed in 
refs.~\cite{Becirevic:2012jf,Datta:2012qk,Tanaka:2012nw,Sakaki:2013bfa,Biancofiore:2013ki,Sakaki:2014sea,Alonso:2015sja,Freytsis:2015qca}.

\label{sec:pheno}

These plots show the different dependencies of the observables on the NP scenarios
considered here which lead to results unambiguously distinct from the SM, even accounting for the theoretical uncertainties.  
Hence, a measurement of the angular observables with enough precision could help to confirm and eventually 
identify the contribution responsible for the enhancements measured in $R_{D^{(*)}}$. The $A^{(*)}_{FB}$ present a mild dependence
on the scalar and current interactions which is not very different from the one of $R_{D^{(*)}}$. On the other hand, the asymmetries are very sensitive to 
the tensor interactions, especially for the $BD^*$ channel where $A^*_{FB}$ changes sign for the values of $\epsilon_T^\tau$ considered here. In Tab.~\ref{tab:results} 
we present the corresponding predictions for the integrated observables $R_{FB}^*$ defined in eq.~(\ref{eq:RFB}), which manifest similar patterns 
to those in Fig.~\ref{fig:diffdecays}.

\begin{table}[h]
\centering
\caption{Numerical results on the observables $R_{D^{(*)}}$ and $R_{FB}^*$ in the SM and in the different
benchmark scenarios of NP (see main text) obtained using the inputs listed in Tab.~\ref{tab:FFparam}. The experimental
averages are taken from ref.~\cite{Amhis:2014hma}. \label{tab:results}}
\begin{tabular}{|c|cc|cc|}
\hline
&$R_D$&$R_{FB}$&$R_{D^*}$&$R^*_{FB}$\\
\hline
SM&0.310(19)&0.0183(9)&0.252(4)&0.0310(7)\\
\hline
Current&0.410&0.0242&0.333&0.0410\\
Scalar&0.400&0.0218&0.315&0.0363\\
Tensor&0.467&0.0151&0.346&$-0.0377$\\
\hline
Expt.&$0.391(41)(28)$&--&$0.322(18)(12)$&--\\
\hline
\end{tabular}
\end{table} 

An important feature of our results concerns the absolute value of the contribution of the forward-backward 
asymmetries to the differential decay rate. Indeed, as shown in Fig.~\ref{fig:diffdecays} and Tab.~\ref{tab:results}, 
the contributions of $A_{FB}^{(*)}$ are typically an order of magnitude smaller than $\Gamma_5^{(*)}$. 
In this sense, the $BD$ mode is specially interesting since the contribution of the normalization decay to the 
forward-backward asymmetry is proportional to $\Gamma_{0+}^I$ in eqs.~(\ref{eq:Gammashel}), which in the SM is suppressed by $m_\ell^2$. 
Therefore, $R_{FB}$ is a very clean observable of the $\bar B\to D\tau^-\bar\nu$ decay, at least regarding the possible pollution of the signal from
the normalization mode. The same argument does not follow for the $BD^*$ mode because, besides $\Gamma_{0+}^I$, $A_{FB}^{*}$ also receives contributions
from decays into transversal $D^*$ which are not suppressed by the light-lepton masses. 
Thus, the separation of signal from background is crucial to exploit the forward-backward asymmetry in this case.

\begin{figure}[h]
\begin{tabular}{cc}
  \includegraphics[width=80mm]{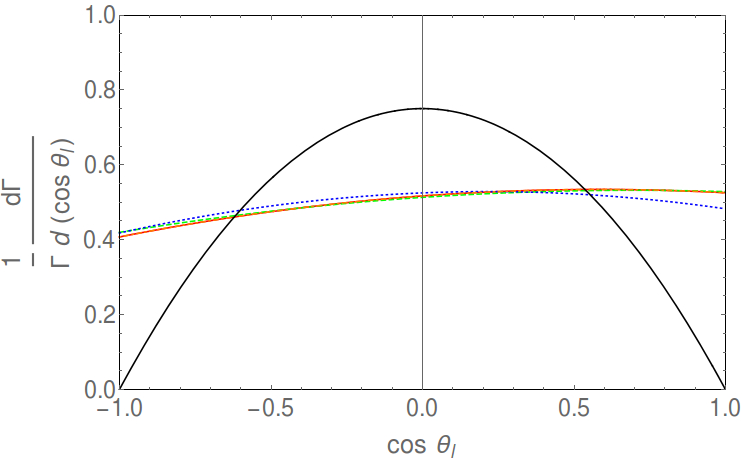}&\includegraphics[width=80mm]{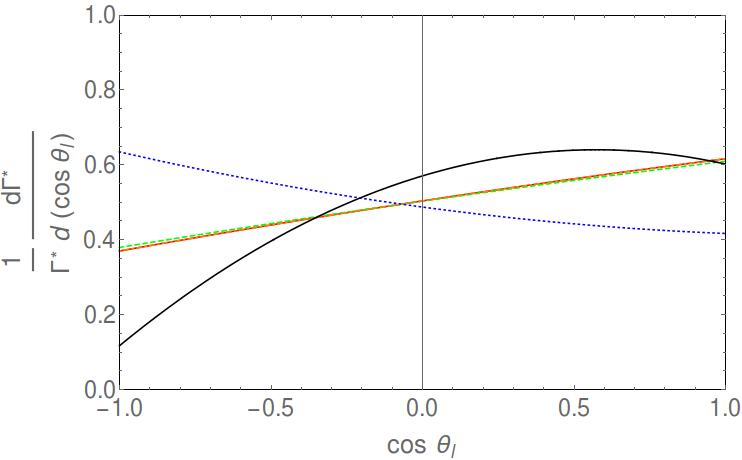}
\end{tabular}
\caption{Angular distribution for the $\bar B\to D^{(*)}\tau^-(\to\ell^-\bar \nu_\ell\nu_\tau)\bar\nu_\tau$ decays,
in the SM and in the different benchmarks of NP (same code as in Fig.~\ref{fig:diffdecays}), compared to the one of 
the normalization mode, $\bar B\to D^{(*)}\ell^-\bar\nu_\ell$.
\label{fig:angdis}}
\end{figure}

The angular analysis can also serve to increase the efficiency of the selection of the signal over the background in these decays.~\footnote{This has 
been pointed out recently and independently in ref.~\cite{Bordone:2016tex} for the $\bar B\to P\tau^-\bar\nu$ decays, with $P=D,\,\pi$.}
As an illustration, we show in Fig.~\ref{fig:angdis}, the angular distributions of the 
$\bar B\to D^{(*)}\tau^-(\to\ell^-\bar\nu_\ell\nu_\tau)\bar\nu_\tau$ and the $\bar B\to D^{(*)}\ell^-\bar\nu$ decays which are quite 
different. The shape of the distribution of $\bar B\to D\ell^-\bar\nu$ is  a consequence of the fact that the only contribution to the rate 
that is not suppressed by $m_\ell^2$ in the SM, enters through $\Gamma_{0-}$ in eqs.~(\ref{eq:Gammashel},~\ref{eq:BDtaunurate}) and behaves 
as $\sim \sin^2\theta_\ell$. In particular, we see another manifestation of the smallness of the forward-backward asymmetry ($A_{FB}^\ell$)
in this case. In contrast to this, the angular distribution of the $\bar B\to D\tau^-(\to\ell^-\bar\nu_\ell\nu_\tau)\bar\nu_\tau$ decay
is approximately flat with a slight tilt produced by $A_{FB}$. On the other hand, the sensitivity of the angular distribution to NP is very small because
the effect of NP in the total rate and $A_{FB}$ for the scenarios considered here are similar, canceling the ratio. 
For the $BD^*$ modes, the sizable $A_{FB}^{*\ell}$, as well as the more complex dependence produced by the contributions to the rates 
stemming from different polarization states of the $D^*$, are visible in the plot. In this case, the effects of the tensor NP scenario
is sizable and modify the slope of the distribution in the SM.

\section{Conclusions}

The discrepancy between $R_{D^{(*)}}$ as measured by three independent experiments and  the standard-model predictions
represents one of the most intriguing anomalies in flavor observables. In order to determine whether this is truly
the manifestation of the long-sought new physics or a misinterpreted background effect, it is crucial to develop new tools 
to analyze all possible observables in the $B\to D^{(*)}\tau\bar\nu$ decay. A strategy for achieving this consists of the analysis of the  
$\bar B\to D^{(*)}\tau^-(\to\ell^-\bar\nu_\ell\nu_\tau)\bar\nu_\tau$ decay as a function of the experimentally-accessible 
variables $q^2$, $E_\ell$ and the angle $\theta_\ell$ between the 3-momentum of the final charged-lepton and the recoiling direction of
the $D^{(*)}$. 

The present work stands as an initial step from the theory side in this direction, providing an analytic formula for the 3-fold 5-body differential decay rate
for both the $BD$ and $BD^*$ modes and including general new physics contributions in the framework of effective field theory.
Besides the $q^2$- and $E_\ell$-spectra of the rates, an angular analysis 
based on $\theta_\ell$ allows to identify new observables independent of the total rates. For instance,
the forward-backward asymmetry captures the contribution to the rate odd under $\theta_\ell\to \pi-\theta_\ell$
which is otherwise invisible integrating over all phase space.
We use this to construct new integrated observables that we call $R_{FB}^{(*)}$. These 
are quite sensitive to the effects of new physics and could provide complementary sources of information
to discriminate among different scenarios. In particular, the asymmetry of the $BD$ mode is very clean in the sense
that any pollution induced by the $B\to D\ell\bar\nu$ decay is negligible. On the other hand,
the asymmetry in the $BD^*$ channel shows a strong sensitivity to tensorial new-physics contributions. 

The angular distribution is not only useful to discriminate among different new-physics scenarios but also to increment the 
efficiency of the selection of the signal over the normalization process. While the dependence on $\cos\theta_\ell$ of the 
$B\to D^{(*)}\ell\bar\nu$ rates presents sizable curvature, the ones for $\bar B\to D^{(*)}\tau^-(\to\ell^-\bar\nu_\ell\nu_\tau)\bar\nu_\tau$ are quite flat,
\textit{viz.} Fig.~\ref{fig:angdis}. 

Our work can be extended to analyze the full kinematic dependence of the rate, for example by looking at the variation of the angular coefficients
$I_i(q^2,\,E_\ell)$ with the lepton energy and transferred momentum, 
or converting the formulae to kinematic variables better suited for a given experiment.
One could also straightforwardly implement
the decay of the $D^*$ in our 5-body formula to obtain the full 6-body differential decay rate. This introduces two new measurable
angles that would lead to a string of new angular observables. Finally, our analytic formulas could help to improve the efficiency 
of the experimental analyses of the data. 

\section{Acknowledgments}

We are especially grateful to Karol Adamczyk and Maria Rozanska for calling our attention to a missing contribution in the first version of this preprint. 
We also would like to thank Marzia Bordone, Ulrik Egede, Ben Grinstein, Gino Isidori, Zoltan Ligeti, Thomas Kuhr, Aneesh Manohar, Mitesh Patel, Vladimir Pascalutsa,
Sascha Turczyk and Danny van Dyk and  for useful discussions. This work was supported in part by DOE grant DE-SC0009919. JMC has received funding 
from the People Programme (Marie Curie Actions) of the European Union's Seventh Framework Programme (FP7/2007-2013) under REA grant agreement
n PIOF-GA-2012-330458. RA thanks the Theory Department at CERN for hospitality during the completion of this work.

\appendix

\section{Analytic formulas for the angular coefficients}
\label{sec:appendix}

\begin{figure}[h]
\begin{tabular}{cc}
  \includegraphics[width=10cm]{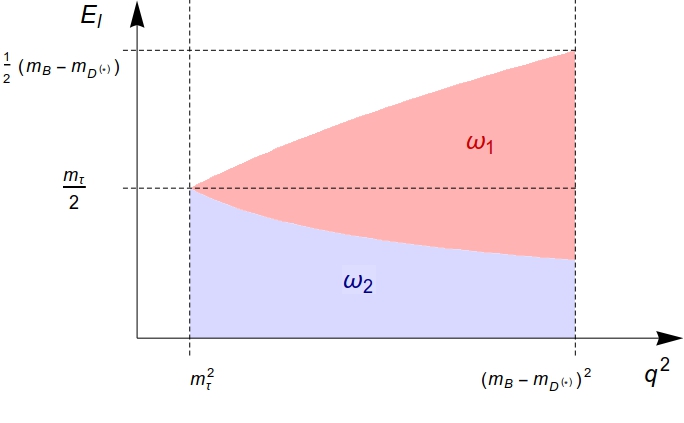}
\end{tabular}
\caption{Phase-space regions of the $I_i(q^2,\,E_\ell)$ angular coefficients.
\label{fig:intreg}}
\end{figure}

The $I_i(q^2,\,E_\ell)$ angular coefficients are piecewise functions with different expressions for the two different  regions of phase space, $\omega_1$ and $\omega_2$, as described in Sec.~\ref{sec:5body} and illustrated in
Fig.~\ref{fig:intreg}. These functions are best presented via the introduction of dimensionless variables
\begin{align}
x^2&=\frac{q^2}{m_\tau^2}\,, & y=\frac{E_\ell}{m_\tau}\,,
\end{align}
and introducing the coefficients of the different powers of $\cos\theta_\tau$ in eq.~(\ref{eq:BDtaunurate}), 
\begin{align}
\Gamma_{-}^{(0)}=&\Gamma_{+-}+\Gamma_{--}+2\Gamma_{0-}\,,\nonumber & \Gamma_{+}^{(0)}=&2\Gamma_{0+}^t+\Gamma_{++}+\Gamma_{-+}\,,\\
\Gamma_{-}^{(1)}=&2\Gamma_{--}-2\Gamma_{+-}\,,\nonumber &\Gamma_{+}^{(1)}=&2\Gamma_{0+}^I\,, \\
\Gamma_{-}^{(2)}=&\Gamma_{+-}+\Gamma_{--}-2\Gamma_{0-}\,, & \Gamma_{+}^{(2)}=&2\Gamma_{0+}^0-\Gamma_{++}-\Gamma_{-+}\,,
\end{align}
or in eq.~(\ref{eq:GammasInt}):
\begin{align}
\mathcal I^{(0)}=2{\rm Re}\left[2\mathcal I_0^I+\mathcal I_++\mathcal I_-\right],\\
\mathcal I^{(1)}=2{\rm Re}\left[2\mathcal I_0+\mathcal I_--\mathcal I_+\right].
\end{align}
For the region $\omega_1$  we have,
\begin{align}
I_0=&\frac{\left(3 x^2-2\right) (x+4 y) (x-2 y)^2}{6 x^2
   \left(x^2-1\right)^2 y^2}\Gamma_{+}^{(0)} +\frac{ \left(2 x^4-3 x^2-16 x
   y^3+12 y^2\right)}{6 x \left(x^2-1\right)^2 y^2}\Gamma_{-}^{(0)}\nonumber \\
   &+\frac{
   \left(20 x^5 y+x^4 \left(40 y^2-6\right)+16 x^3 y \left(5
   y^2-4\right)+x^2 \left(15-72 y^2\right)-4 x y \left(8 y^2-5\right)+20
   y^2\right) (x-2 y)^2}{120 x \left(x^2-1\right)^4 y^4}\Gamma_{-}^{(2)}\nonumber \\&
   +\frac{ \left(40 x^5 y+5 x^4 \left(16 y^2-3\right)-50 x^3 y+x^2 \left(6-80
   y^2\right)+16 x y+24 y^2\right) (x-2 y)^3}{120 x^2 \left(x^2-1\right)^4
   y^4}\Gamma_{+}^{(2)}\\
   &+\frac{-240 x^5 y^4+9 x^5+32 \left(10 x^4-5 x^2+1\right) y^5-30 \left(x^2+1\right) x^4 y+20 
   \left(x^4+4 x^2+1\right) x^3 y^2}{120 x \left(x^2-1\right)^4 y^4}\mathcal I^{(1)}\,,
\end{align}
\begin{align}
I_1=&\frac{ \left(-2 x^4+x^2+4 \left(3 x^4-3 x^2+1\right)
   y^2+\left(3 x^4-5 x^2+2\right) x y\right) (x-2 y)^2}{6 x^2
   \left(x^2-1\right)^3 y^3}\Gamma_{+}^{(1)}\nonumber \\&+\frac{ \left(2 x^6 y-x^5-3
   x^4 y+x^3 \left(2-16 y^4\right)+x^2 y \left(20 y^2-3\right)-4
   y^3\right)}{6 x \left(x^2-1\right)^3 y^3}\Gamma_{-}^{(1)}\\
   &-\frac{(x-2 y)^2 \left(2 x^3 y+x^2 
   \left(8 y^2-1\right)-2 x y-4 y^2\right)}{6 x \left(x^2-1\right)^3 y^3}\mathcal I^{(0)}\,,
\end{align}
\begin{align}
I_2=&\frac{1}{120
   \left(x^2-1\right)^4 y^4} \Big[720 x^3 y^4-64 \left(5
   \left(x^4+x^2\right)-1\right) y^5-60 x^2 \left(x^4-2 x^2-2\right) y+9 x^3
   \left(2 x^2-5\right)\nonumber\\
   &+20 x \left(2 x^6-x^4-16 x^2-3\right) y^2\Big]\Gamma_{-}^{(2)}
   \nonumber
   +\frac{1}{120 x^2 \left(x^2-1\right)^4 y^4} \Big[-720 x^7 y^4+9
   \left(5 x^2-2\right) x^5\\&-60 \left(2 \left(x^4+x^2\right)-1\right) x^4
   y+64 \left(5 \left(3 x^6-2 x^4+x^2\right)-1\right) y^5+20 \left(3 x^6+16
   x^4+x^2-2\right) x^3 y^2\Big]\Gamma_{+}^{(2)}\\
   &+\frac{240 x^5 y^4-9 x^5-32 \left(10 x^4-5 x^2+1\right) y^5+30 \left(x^2+1\right) x^4 y-20 
   \left(x^4+4 x^2+1\right) x^3 y^2}{40 x \left(x^2-1\right)^4 y^4}\mathcal I^{(1)}\,,
\end{align}
and for the region $\omega_2$,
\begin{align}
I_0=&-\frac{2  \left(2 x^2+1\right) (4 x y-3)}{3 x}\Gamma_{-}^{(0)}+\frac{2
    \left(x^2+2\right) (3 x-4 y)}{3 x^2}\Gamma_{+}^{(0)}\nonumber \\&
   +\frac{2}{15}
    \left(-12 x^2 y+10 x+\frac{5}{x}-8
   y\right)\Gamma_{-}^{(2)}+\frac{ \left(10 x \left(x^2+2\right)-8 \left(2
   x^2+3\right) y\right)}{15 x^2}\Gamma_{+}^{(2)}-\frac{4 \left(x^2-1\right) y}{15 x}\mathcal I^{(1)}\,,
\end{align}
\begin{align}
I_1=&\frac{\left(8 x^3 y-4 x^2+2\right)}{3 x}\Gamma_{-}^{(1)} -\frac{2
    \left(x^3-2 x+4 y\right)}{3 x^2}\Gamma_{+}^{(1)}+\frac{4}{3} \left(-2 x y-\frac{2 y}{x}+1\right)\mathcal I^{(0)}\,,
\end{align}
\begin{align}
I_2=\frac{8  \left(x^2-1\right) y}{15 x^2}\Gamma_{+}^{(2)}-\frac{8}{15} \left(x^2-1\right) y   \Gamma_{-}^{(2)}
+\frac{4 \left(x^2-1\right) y}{5 x}\mathcal I^{(1)}\,.
\end{align}

\bibliography{BtoDtaunu.bib}
\end{document}